\documentclass[11pt]{article}

\usepackage[english]{babel}
\usepackage{float}
\usepackage{amsmath,array,amssymb}
\usepackage{graphicx}
\usepackage{braket}
\usepackage{siunitx,bigints}
\usepackage{multirow}
\usepackage{amsfonts}
\usepackage{setspace}
\usepackage[infoshow,debugshow]{tabularx}
\usepackage{braket}
\usepackage{lineno}
\usepackage{color}
\usepackage{placeins}
\usepackage{multirow}
\usepackage{booktabs}
\usepackage{threeparttable}
\usepackage{adjustbox}
\usepackage{xr-hyper}

\usepackage[
    backend=biber,
    style=numeric-comp, % numerica e compatta
    sorting=none,
    doi=false,
    url=false,
    eprint=false,
    isbn=false
]{biblatex}

\AtEveryBibitem{%
  \clearfield{doi}%
  \clearfield{url}%
  \clearfield{eprint}%
  \clearfield{isbn}%
}

\usepackage{hyphenat}
\usepackage[letterpaper,top=2cm,bottom=2cm,left=3cm,right=3cm,marginparwidth=1.75cm]{geometry}

\usepackage{amsmath}
\usepackage{graphicx}
\usepackage[colorlinks=true, allcolors=blue]{hyperref}
\begin{document}
	
% --- Titolo compatto stile IOP ---
\title{\bfseries Optical investigation of ultra-slow spin relaxation in $^{171}$Yb$^{3+}$:Y$_2$SiO$_5$ single crystals}

\author{
\parbox{\textwidth}{\centering
F. Chiossi$^{1}$\thanks{Present address: Department of Physics and Astronomy, University of Padova, Padova, 35131, Italy. Email: federico.chiossi@chimieparistech.psl.eu}, 
A. Tiranov$^{1}$, 
L. Nicolas$^{2}$, 
D. Serrano$^{1}$, 
F. Montjovet-Basset$^{1}$, \\
E. Lafitte-Houssat$^{1,4}$, 
A. Ferrier$^{1,4}$, 
S. Welinski$^{3}$, 
L. Morvan$^{3}$, 
P. Berger$^{3}$, 
A. Tallaire$^{1}$, \\
M. Afzelius$^{2}$, 
P. Goldner$^{1}$\thanks{Email: philippe.goldner@chimieparistech.psl.eu}
}}

\date{
\small
$^{1}$Chimie ParisTech, PSL University, CNRS, Institut de Recherche de Chimie, Paris, 75005, France\\
$^{2}$Department of Applied Physics, University of Geneva, Geneva 4, Switzerland\\
$^{3}$Thales Research and Technology, 91767 Palaiseau, France\\
$^{4}$Faculté des Sciences et Ingénierie, Sorbonne Université, UFR 933, 75005 Paris, France
}

\maketitle

\begin{abstract}
\noindent

We present a comprehensive study of spin relaxation dynamics at cryogenic temperatures in a rare-earth-doped crystal used for quantum memory applications: $^{171}$Yb:Y$_2$SiO$_5$. Spin relaxation is indeed a major limiting factor for both the efficiency and storage time of quantum memory protocols based on atomic frequency combs in rare-earth materials. The relaxation dynamics among the four ground-state hyperfine levels were simultaneously investigated by optically perturbing the spin population distribution and monitoring its return to thermal equilibrium through optical absorption spectroscopy. By applying different types of perturbations, we were also able to distinguish between two types of relaxation processes, induced by spin-phonon and spin-spin interactions.

Below 1\,K, we observed that the re-thermalization of the Yb$^{3+}$ ion population takes several hours, driven solely by direct phonon absorption or emission. However, the effective lifetime of individual spin states is much shorter — on the order of several seconds in low-doped (2\,ppm) samples and of milliseconds in 10\,ppm samples — due to spin-spin interactions. These findings provide valuable guidelines for optimizing doping levels and operating temperatures in rare-earth-doped crystals for quantum applications. Notably, they suggest that atomic frequency combs with lifetimes of several hours could be realized using $^{171}$Yb:Y$_2$SiO$_5$ crystals with slightly less than 2\,ppm doping and operating near 1\,K.

\end{abstract}

%\ioptwocol

\section{Introduction}

One of the main challenges in realizing long-distance quantum communication networks lies in overcoming the signal attenuation in optical fibers\,\cite{chen2021twin, pittaluga2025long}. Since quantum signals cannot be amplified, the only viable solution appears to be the implementation of quantum repeaters\,\cite{briegel1998quantum,sangouard2011quantum}. However, their realization is currently limited by the lack of efficient and reliable quantum memories\,\cite{simon2007quantum}. Among the several approaches to storing quantum information, the atomic frequency comb (AFC) based protocol stands out as one of the most promising thanks to its capability of working at a single photon level with multiplexing in frequency and space, and its potential to achieve high efficiency and fidelity\,\cite{deriedmattenSolidstateLightMatter2008,zhouPhotonicIntegratedQuantum2023,durantiEfficientCavityassistedStorage2024,tittel2025quantum}.
In the AFC protocol, the absorption profile of an optical medium is finely tailored into a periodic, comb-like structure with the "teeth" spaced by multiples of a fixed frequency $\Delta$. When excited by broadband photons, the atoms in these teeth experience a rephasing at times multiple of $2\pi/\Delta$ and the stored excitation is coherently re-emitted\,\cite{afzeliusMultimodeQuantumMemory2009,bonarota2010efficiency}. Tailoring the absorption profile is generally accomplished by selectively moving absorbing ions to another level via optical excitation. Such a technique is known as spectral hole burning and requires optical levels or different spin levels in the ground state with a long lifetime.
In the AFC protocol, however, the readout is not on demand and the storage time is limited to the optical coherence time. To overcome this limitation, a variant of the AFC protocol, known as the spin-wave AFC protocol was proposed\,\cite{afzelius2010demonstration}. In this variant, the excitation stored in the optical comb is coherently transferred to a ground spin state, where quantum states can be stored for longer times, and subsequently re-transferred to the optical comb for the readout.

The rare-earth-doped crystals (REC) are ideal candidates for implementing spin-wave AFC quantum storage thanks to their long coherence times ($T_2$) and large absorption bands that can be easily tailored\,\cite{Goldner2015,thielRareearthdopedMaterialsApplications2011,zhong2019emerging,guo2023rare,ortu2022multimode}. After the first demonstration of the AFC optical storage in 2008\,\cite{deriedmattenSolidstateLightMatter2008}, storage times exceeding hundreds of milliseconds and up to one hour followed using non-Kramer RE ions several years later\,\cite{holzapfel2020optical,maOnehourCoherentOptical2021}. However, in these experiments, storage efficiency is still restricted to a few percent or below for delays longer than 1\,ms\,\cite{ortu2022storage,ortu2022multimode,laplane2015multiplexed,maOnehourCoherentOptical2021}. The operational bandwidth, directly related to the width of the frequency comb, is another factor to consider. In non-Kramers ions, it is generally tens of MHz at most, set by the small nuclear hyperfine splittings. Kramers ions, featuring much larger splitting\,\cite{cruzeiro2018efficient,businger2022non,vivoli2013high}, can reach up to a few GHz bandwidth but the maximum storage duration is generally shorter as their quantum states are more sensitive to magnetic fluctuations\,\cite{macfarlane2002high}. In this context, $^{171}$Yb:Y$_2$SiO$_5$ ($^{171}$Yb:YSO) emerged as a promising platform for quantum storage since it incorporates the large hyperfine splitting typical of the Kramer ions and, thanks to clock transitions, millisecond-long optical and spin coherence times at zero magnetic field, comparable to those of the non-Kramer ions\,\cite{Ortu2018,Tiranov2018,chiossi2024optical}. %Using a crystal doped at 5\,ppm, an AFC-based quantum memory working at 978\,nm, with a bandwidth of 100\,MHz and a storage time of 25 microseconds was realized. More recently, the implementation of the spin-wave AFC protocol extended the storage duration up to a few milliseconds (unpublished??).
%In the previous work, we showed that the doping concentration plays an important role in the Yb:YSO properties, notably coherence time and spectral diffusion. The spectral diffusion is  
A key limitation to the storage efficiency and duration in AFC-based quantum memory arises from spin-flip processes that modify the optically prepared spin distribution, leading to the degradation of the frequency comb. Moreover, several hundreds of milliseconds are required to re-create the frequency comb, during which no photons can be stored, effectively reducing the duty cycle of the quantum memory. Understanding the mechanisms behind spin-flip processes and finding ways to reduce their rate or mitigate their effects is therefore crucial for improving memory performance. The spin relaxation of $^{171}$Yb ions in YSO has been previously investigated, but only up to a time scale of a few milliseconds, primarily to assess its impact on the Yb coherence time\,\cite{chiossi2024optical}. 

In this work, we extend the study of spin relaxation dynamics in $^{171}$Yb samples over timescales ranging from milliseconds to hours, distinguishing between processes driven by spin-lattice and spin-spin interactions. We investigate two $^{171}$Yb:YSO samples with doping concentrations of 10\,ppm and 2\,ppm, across a temperature range from 6\,K to 50\,mK. The spin relaxation rates are determined by optically perturbing the ground-state spin population distribution and monitoring its recovery through variations in the recorded optical absorption spectrum. Compared to the microwave excitation, this approach has the advantage of directly probing the population dynamics of all four hyperfine levels in $^{171}$Yb ground state, providing a more comprehensive insight into the underlying spin relaxation mechanisms.

After a theoretical description of the spin‐flip processes (Sec.\,\ref{sec:theory}), we present our experimental results in Sec.\,\ref{sec:results}, which is divided into two subsections. In the first one (Sec.\,\ref{sec:recovery_measurements}), we focus exclusively on spin–phonon interactions, whereas in the second we also incorporate spin–spin processes. In both cases, we employ analytical expressions derived by solving coupled rate‐equation systems that model the spin dynamics; these equation systems are detailed in the Appendix for easier reading. Finally, we discuss the implications of our findings in the context of theoretical predictions and previous experimental studies.

%
%While the SLR is strongly temperature dependent, the spin FF rate depends on the average distance between resonant ions and thus the doping concentration. Our study on these two processes can help define the temperature range in which the comb degradation time is mainly limited by the spin flip-flop process and assess the time that can be reached by decreasing the doping concentration. 
%The FF and SLR in Yb:YSO are theoretically described in Sec.\,\ref{sec:theory}. In the next two sections, the spin relaxation is investigated in two Yb:YSO crystals, doped at 2 and 10\,ppm, through measurements of the spin thermalization time and spectral hole decay, for temperatures between 50\,mK and 6\,K. 
%The study of these two kinds of spin relaxation is thus necessary to assess the comb degradation time for different temperatures and crystal doping concentrations.
%and the population recovery is solely accomplished by the SLR processes. These measurements thus allow us to derive the ultimate limit of the frequency comb lifetime 

\section{Spin relaxation theoretical model}
\label{sec:theory}
Spin relaxation processes of a Kramers ion in a solid-state material, such as Yb$^{3+}$ in YSO, is primarily caused by interactions with
lattice phonons and by a direct exchange of spin levels between nearby ions. These mechanisms are commonly referred to as spin-lattice relaxation (SLR) and spin flip-flop (FF) processes, respectively. When a Yb ion moves to a different spin state, its optical absorption frequencies shift accordingly. Nearby ions also undergo slight frequency shifts due to magnetic dipole interactions, an effect known as spectral diffusion\,\cite{Bottger2006}. Consequently, spin relaxation can alter the precisely prepared spin state distribution and lead to the degradation of the AFC optical absorption profile. %More specifically, ions that initially contribute to absorption in one of the AFC teeth may move to a different spin level, resulting in absorption in one of the AFC holes, or vice versa. 
The next subsections review the key features of SLR and FF processes. For the purposes of this work, we restrict our analysis to the case of zero applied magnetic field, for which long optical and spin coherence time have been observed, as mentioned above. A third subsection then presents a concise description of the $^{171}$Yb ground-state hyperfine structure in YSO, along with any additional relevant information.

\subsection{Spin flip-flop relaxation and analysis}
\label{sec:spin-spin}
In FF processes, two resonant ions exchange their spin states through magnetic dipole-dipole interaction. The rate of this process can be calculated based on the Fermi Golden Rule,\cite{car2019optical}, and we summarize the main results here. The FF rate for a specific ion $i$, due to the magnetic interaction with the surrounding ions in state $j$, can be expressed as:
\begin{equation}
    F_{ij} = \frac{2\pi}{\hbar} \left( \frac{\mu_0 \mu_B^2}{4\pi} \right)^2  \frac{\Xi \braket{r_{ij}^{-6}}}{\hbar \Gamma_{ihn,s}} 
\end{equation}
where $\mu_0$ and $\mu_B$ denote the permeability of free space and the Bohr magneton, $\Xi$ is a dimensionless ion-ion coupling factor, $\Gamma_{ihn,s}$ is the inhomogeneous linewidth of the $i\leftrightarrow j$ spin transition and $\braket{r_{ij}^{-6}}$ the average inverse sixth power of the distance between the ion in state $i$ and the surrounding ions in state $j$\,\cite{car2019optical}. The rate of this process depends on the ion concentration and the instantaneous spin population distribution within the crystal, with temperature affecting it only indirectly through modifications of the spin distribution. It should also be noted that while the state exchange alters the individual ion state populations, it does not affect the overall population distribution, meaning that the fraction of ions occupying each energy level remains unchanged.

\subsection{Spin-lattice relaxation}
\label{sec:spin-lattice}
SLR processes involve the interaction of ions with phonons, generally driving out-of-equilibrium populations toward the Boltzmann equilibrium distribution. These processes are strongly influenced by the phonon state population in the lattice and, consequently, by the sample temperature T.

At low temperatures, typically below $\sim$1 K, the spin-lattice relaxation is governed by the first-order mechanism known as the direct process. In a two-level system, in which the spin level $i$ lies below $j$, the upward direct process $S^{\uparrow}_\text{d,ij}$ can occur only through the resonant absorption of one phonon. Conversely, the downward direct process $S^{\downarrow}_\text{d,ji}$ from j to i can be triggered either by spontaneous or stimulated phonon emission.  By assuming a Debye model for the phonon distribution, it is possible to show that the two direct processes rate can be expressed as\,\cite{abragam2012electron}:
\begin{equation}
\left \{
\begin{array}{rcl}
S^{\uparrow}_\text{d,ij} &=&  \dfrac{3\omega_{ij}^3|\bra{i}\epsilon V\ket{j}|^2}{2\pi \hbar \rho v^5}  \dfrac{1}{\exp(\hbar \omega_{ij} / k_b T)-1} =   \dfrac{A_{ij}(\hbar \omega_{ij})^3}{\exp(\hbar \omega_{ij} / k_b T)-1}  \\ \\
S^{\downarrow}_\text{d,ji} &=& \dfrac{3\omega_{ij}^3|\bra{i}\epsilon V\ket{j}|^2}{2\pi \hbar \rho v^5}  \dfrac{ \exp(\hbar \omega_{ij} / k_b T)}{\exp(\hbar \omega_{ij} / k_b T)-1} =   \dfrac{A_{ij}(\hbar \omega_{ij})^3\exp(\hbar \omega_{ij}/ k_b T)}{\exp(\hbar \omega_{ij} / k_b T)-1}  \\ 
\end{array}
\right.
\label{Eq:direct_rate}
\end{equation}
where $k_b$ is the Boltzmann factor, $\hbar \omega_{ij}$ is the energy of the phonon, $\rho$ and $v$ the density and the speed of sound in the material, $\epsilon V$ is the first-order variation of the crystalline electric field operator generated by the lattice strain. Note that, in the final step, the constants and the matrix element have been incorporated into the $A_{ij}$ parameters, effectively isolating the dependencies on the transition frequency and the sample temperature.

For $\hbar \omega \gg k_b T$, the low phonon density suppresses the phonon absorption and stimulated emission processes, leading to $S^{\uparrow}_\text{d,ij} = 0$ and $S^{\downarrow}_\text{d,ji} = A_{ij}(\hbar \omega_{ij})^3$. For $\hbar \omega \ll k_b T$, the rates can be approximated to the first order as $S^{\uparrow}_\text{d,ij} = S^{\downarrow}_\text{d,ji} = A_{ij}(\hbar \omega_{ij})^2k_bT$, exhibiting a linear dependence on the temperature. The applied model assumes that the phonon occupation number reliably follows the Bose-Einstein statistics. However, the density for low-energy phonons may be lower as the presence of defects and the finite size of the crystal limit the phonon mean free path and thus the number of phonon modes\,\cite{kittel2004phonons,liu2002restricted}. Additionally, at ultra-low temperatures, phonons generated during spin-lattice relaxation may not dissipate efficiently into the thermal bath, leading to a significant out-of-equilibrium phonon population capable of directly re-exciting the ions. These deviations from the Debye model, commonly referred to as phonon-bottleneck effects, reduce the rate at which the spin population reaches thermal equilibrium. When these effects become particularly significant, the spin-lattice relaxation rate is observed and expected to follow a quadratic temperature dependence, deviating from the linear relationship predicted in their absence\,\cite{abragam2012electron,scott1962spin,budoyo2018phonon}.

For temperatures above $\sim$1\,K, higher-order spin-lattice relaxation (SLR) processes must be taken into account\,\cite{kurkin1980epr}. These include the Raman and Orbach processes, both of which involve the absorption of one phonon from the lattice and the simultaneous emission of another with a different energy. In the Orbach process, however, the absorbed and the emitted phonon are each resonant with a distinct ion transition. The SLR rate for a Kramers ion is thus generally approximated with a combination of three terms\,\cite{abragam2012electron}:
\begin{equation}
S = S_{d} + c{T^9} + \frac{\alpha_O}{\exp(\Delta /k_b T)-1},
\label{Eq:SLR_total}
\end{equation}
where $S_{d}$ represents the rate of the upward or downward direct process, $c$ is the coefficient for the Raman process, $\alpha_O$ characterizes the strength of Orbach processes, and $\Delta$ is the energy separation between the occupied energy level and the nearest Stark crystal field level.%the smallest transition energy between Stark levels.  

\subsection{Spin relaxation model in Yb-171 in YSO}
\label{sec:spin-spin2}
In this work, we focus exclusively on $^{171}$Yb ions at site II of the YSO matrix, owing to their stronger optical absorption and their use in AFC‐based quantum storage\,\cite{businger2022non}. The ground‐state Stark manifold is split into four singlet hyperfine sublevels (spin levels) with quenched magnetic moments \cite{Ortu2018}. The splittings in order of increasing level energy are 528 MHz, 1.842 GHz, and 655 MHz (see Fig.\,\ref{fig:Figure1}c)\,\cite{Welinski2016,Tiranov2018}.

The first excited Stark level lies at 234 cm$^{-1}$ above the ground state, a gap large enough to suppress the Orbach relaxation channel, at least at temperatures below 20\,K\,\cite{chiossi2024optical}. In principle, each pair of spin levels $(i,j)$ (with $i,j\in{1,2,3,4}$ and $i>j$) supports distinct direct and Raman relaxation rates, all contributing to the overall spin‐population dynamics. Unfortunately, the low symmetry of the crystal field ($C_1$) renders the coupling constants $A_{ij}$ (direct processes) and $c_{ij}$ (Raman processes) exceedingly difficult to compute from first principles, so we treat them as free parameters to be extracted experimentally. Previous short‐time spin‐dynamics measurements have shown that the $c_{ij}$ values are of the same order of magnitude\,\cite{chiossi2024optical}.

By contrast, the flip–flop (FF) rates $F_{ij}$ can be calculated analytically \cite{Welinski2020}. Owing to the symmetry of the $^{171}$Yb wavefunctions in YSO due to the high crystal anisotropy, each FF rate for a given level pair equals that of its symmetric counterpart. For site II ions in a 10\,ppm‐doped crystal, $F_{12} = F_{34}= 390\,$s$^{-1}$, $F_{13} = F_{24}= 1.5\,$s$^{-1}$ and $F_{14} = F_{23}= 0.03\,$s$^{-1}$ were calculated. In Ref.\,\cite{Welinski2020, chiossi2024optical}, the $F_{34}$ was measured using spectral hole burning, yielding a value of approximately $\sim$500\,s$^{-1}$, which is in reasonable agreement with theoretical predictions.

\section{Spin relaxation measurements and analysis}
\label{sec:results}
The spin relaxation was investigated in two Yb:YSO crystals doped at 2 and 10\,ppm for temperatures between 50\,mK and 6\,K by perturbing the $^{171}$Yb ions population distribution and observing how it returned to equilibrium. We devoted a particular attention to the spin relaxation of the 4g level since it is the lower level of the optical transition used for most AFC-based quantum memory experiments to date.  

The perturbation was accomplished by pumping the ions 
from the ground $^{2}F_\text{7/2}(0)$ to the optical $^{2}F_\text{5/2}(0)$ hyperfine levels  (around 978.85 nm in vac.),
and letting the ions freely relax to the ground spin levels. The experimental setup, which includes a narrow tunable diode laser (Toptica, DL pro) and a dilution refrigerator (Bluefors SD), was described in Ref.\,\cite{chiossi2024optical}. The same laser setup was also used to record the evolution of the optical absorption spectrum and derive the dynamics of the level population after the perturbation. During all measurements, the laser beam propagated along the b axis with its polarization set along the D2 crystal axis. 

Two distinct laser pulse sequences were employed to manipulate the spin-level populations of Yb ions. In the first approach, most of the Yb ions initially occupying the 1g and 2g levels were optically pumped to the 4g level. This method allowed us to investigate population relaxation (Section\,\ref{sec:recovery_measurements}). By analyzing data separately below 1.5\,K and above 1.7\,K, we independently estimated the direct (\ref{sec:model1500}) and Raman (\ref{sec:model1700}) coupling rates associated with SLR processes. As previously mentioned, flip-flop (FF) processes do not impact the overall population relaxations. In the second approach, known as spectral hole burning (SHB), a subset of ions with 4g$\leftrightarrow$1e transition frequencies falling within a few MHz excitation bandwidth out of $\sim$1\,GHz-broad inhomogeneous linewidth were transferred to the other ground state levels, leaving the majority of ions unperturbed. Focusing on this specific subset enabled us to estimate the rate of the FF process in addition to the Raman process, as the FF process redistributes the deficit of ions in that optical subset across all others (Section\,\ref{sec:SHB}).

\subsection{Population recovery}
\label{sec:recovery_measurements}
\begin{figure}[h!]
	\includegraphics[width=1\linewidth]{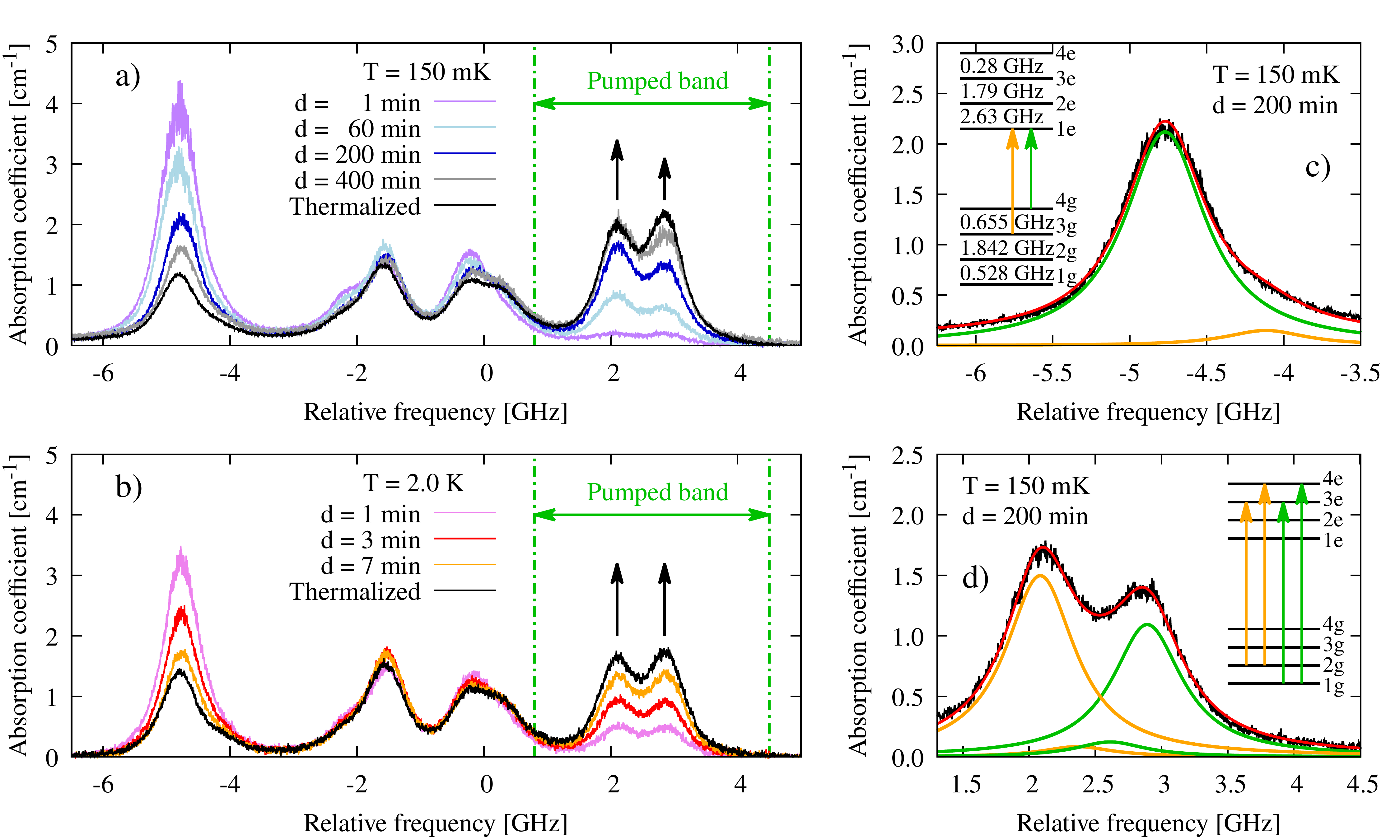}
	\caption{Evolution of the absorption spectrum of Yb doped YSO at 10\,ppm occupying site II recorded at the crystal temperature of 150\,mK (a) and 2.0\,K (b) and at varying delays after the initialization pulses. The green arrows indicate the spectrum region pumped by the initialization pulses. The black arrows highlight the different recovery behavior of the two absorption peaks at the two temperatures. Portion of the absorption spectrum used to estimate the relative population of levels 4g and 3g (c), and of 1g and 2g (d). The red curves represent the best-fit results using four  Lorenztian functions (d) or two Lorentzian functions, and an offset (c). The profile of the single Lorenztians is drawn in green and orange. The absorption spectrum of the 2\,ppm sample can be found in Ref.\,\cite{lafitte-houssatOpticalHomogeneousInhomogeneous2022}}
	\label{fig:Figure1}
\end{figure}

The absorption spectrum of the Yb:YSO crystal comprises 16 partially-resolved transitions (Fig.\,\ref{fig:Figure1}). Strong laser pulses scanning a spectral range that encompasses four transitions starting from the 1g and 2g, and partially a transition from the 3g level (see the green double-headed arrow in Fig.\,\ref{fig:Figure1}a and Fig.\,\ref{fig:Figure1}b) were used to drive more than 65\% of the total Yb ion population to the 4g level. The population in the 3g level was only slightly affected by the excitation, while almost 10\% of the total ions remained in the 1g and 2g levels, distributed roughly equally. Following this initialization, several weak laser pulses scanning a broader band were sent to the crystal at regular delays to record the evolution of the entire absorption spectrum which is directly linked to the ground state spin population distribution. Representative spectra recorded at 150\,mK and 2\,K are presented in Fig.\,\ref{fig:Figure1} (a) and (b) for reference. The absorption profile at the lowest energy (Fig.\,\ref{fig:Figure1}c) containing solely transitions from the 4g and 3g levels was fitted with two Lorentzian functions and an offset to extract the relative population of the two levels. Similarly, the absorption spectrum at the highest energy (Fig.\,\ref{fig:Figure1}d) was analyzed to infer the population at the 1g and 2g levels. Four Lorentzians were used to fit the absorption profile. The width of the Lorentzian functions was fixed to the same value (0.64\,GHz for the 10\,ppm sample and 1.0\,GHz for the 2\,ppm one) for all datasets, as well as the relative position of their center according to the hyperfine level separations reported in Ref.\,\cite{Tiranov2018}.  

\begin{figure}[h!]
	\includegraphics[width=1\linewidth]{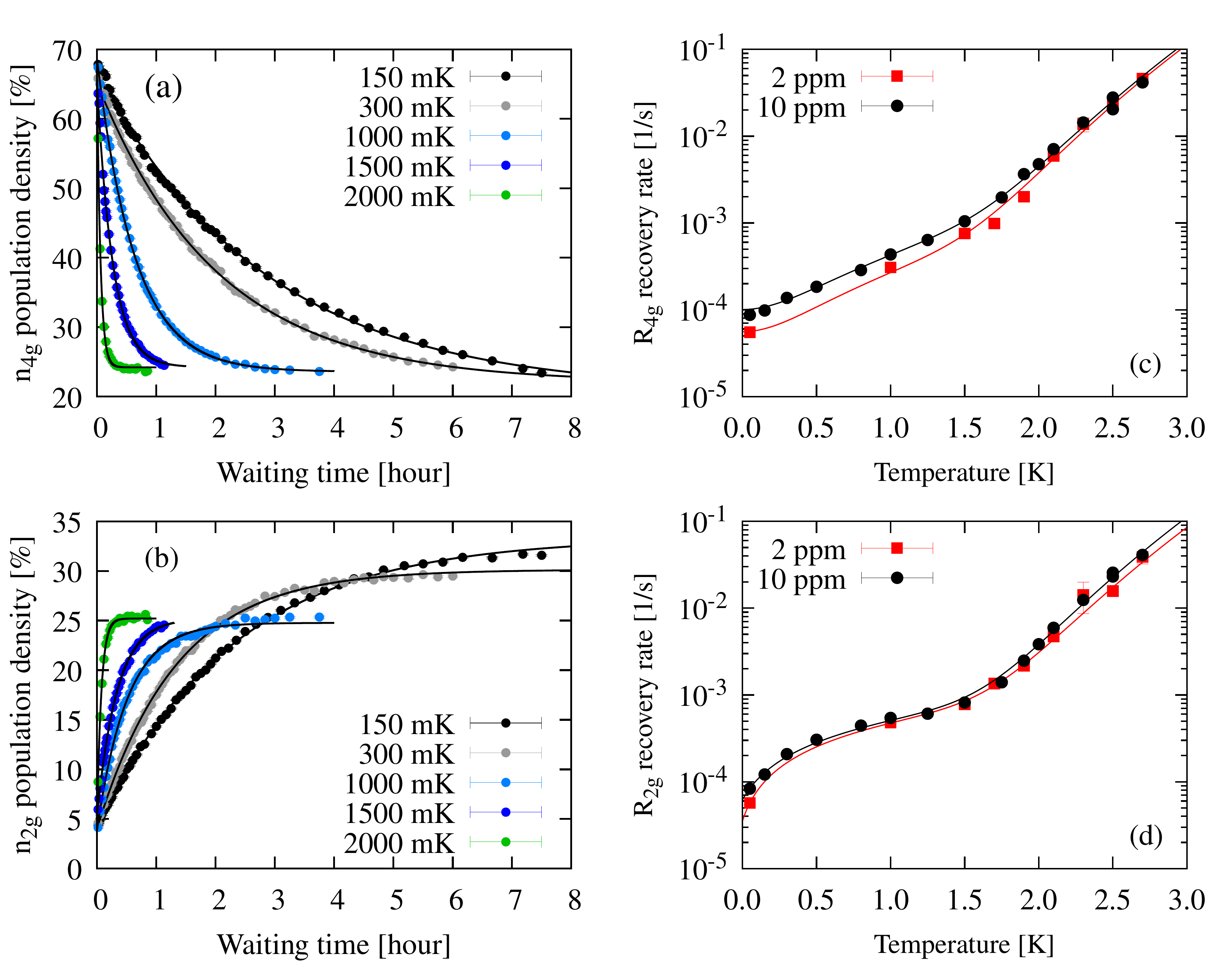}
	\caption{Population density recovery of 4g (a) and 2g (b) levels for five temperatures recorded in the 10\,ppm sample. The black lines are exponential fits. Population recovery rate estimated through the exponential fits for the 4g (c) and 2g (d) population densities. In black are the results for the 10\,ppm sample and in red for the 2\,ppm one. }
	\label{fig:Figure2}
\end{figure}

The areas of the Lorentzian fits for each delay and temperature up to 2.7\,K were converted into population densities. These two quantities are proportional and the normalizing factor was calculated starting from the Lorentzian area estimated from the absorption spectrum at 2.0\,K at equilibrium, assuming that the level population densities ($n_\text{1g},n_\text{2g},n_\text{3g},n_\text{4g}$) at this temperature follow the Boltzmann distribution. The obtained $n_\text{4g}$ and $n_\text{2g}$ for the Yb:YSO 10\,ppm sample are shown in Fig.\,\ref{fig:Figure2} a and b, for several temperatures between 150\,mK and 2.0\,K, measured up to 8 hours after the initialization pulses. Since $n_\text{1g}$ is similar to $n_\text{2g}$, it is not displayed. Data concerning $n_\text{3g}$ exhibit minimal temporal variation and are subject to significant uncertainties. Their analysis is thus not included in the subsequent discussion (see Fig.\,\ref{fig:Direct} in sec.\,\ref{sec:model1500}).

%The data relating to $n_\text{3g}$ show small variation over time and its analysis is also omitted in the following discussion since is affected by large uncertainties (see Fig.\,\ref{fig:Direct} for more details). 

As depicted in Fig.\,\ref{fig:Figure2} a and b, the initialization pulses resulted in more than 65\% of the Yb ions occupying the 4g level, leaving only a few percent in the 1g and 2g levels. The population of each level then returned to equilibrium, following a trend well approximated by the function:
\begin{equation}
n_\text{ig}(t) = n_\text{ig}^{eq} + (N_\text{ig}-n_\text{ig}^{eq})\exp(-R_\text{ig}t)
\end{equation}
where $N_\text{ig}$ (with i = 1, 2, 3 and 4) is the perturbation induced by optical pumping on the level $ig$ at the temperature $T$ and $R_\text{ig}$ is the recovery rate of the same level population. The equilibrium population densities, $n_\text{ig}^{eq}$, align with the expected values for a Boltzmann distribution at the crystal temperature only above 500\,mK. Below this temperature, a discrepancy emerges between the crystal temperature (measured at the dilution base) and the effective spin temperature, which becomes more pronounced for lower temperatures. At the dilution refrigerator base temperature of 50\,mK, the population densities correspond to an effective spin temperature between 130 and 180\,mK. This discrepancy was attributed to thermal radiation from the 4\,K inner thermal shield of the dilution refrigerator and the weak thermal coupling between the crystal and the dilution unit base.

%This effect is attributed to a weak coupling between the spins and the crystal's thermal bath, as well as residual radiative coupling of the spins with the inner thermal shield of the dilution refrigerator. %Because of this effect, the exponential fitting was performed using three free parameters: $n_\text{ig}^{eq}$, $N_\text{ig}$ and $R_\text{ig}$.

As the temperature decreases, the population recovery rate slows significantly, requiring several hours to reach equilibrium below 1\,K. The recovery rates $R_\text{4g}$ and $R_\text{2g}$, as functions of temperature, are shown in black in Fig.\,\ref{fig:Figure2} (c),(d). Similar measurements and analyses were performed using a Yb:YSO crystal doped at 2\,ppm, with the corresponding recovery rates added to the figure in red. Larger statistical errors were observed for this sample due to its lower absorption coefficient and larger inhomogeneous linewidth. The rates related to $n_\text{1g}$ exhibit behaviors similar to those of $n_\text{2g}$ and are not shown in this figure but are reported in Fig.\,\ref{fig:Figure4} for comparison with the spectral hole relaxation rates.

%Compared to the rates derived from the 10\,ppm sample, the ones of the lower concentrated sample are quite close at temperatures above 2\,K but slightly smaller at lower temperatures.

At temperatures above 1.8\,K, all recovery rates increase proportionally to the ninth power of the temperature and converge to similar values. Below this threshold, their temperature dependencies diverge: the rate associated with $n_\text{4g}$ decreases quadratically with temperature, while those related to $n_\text{1g}$ and $n_\text{2g}$ exhibit a linear decrease. Consequently, $R_\text{4g}$ data were fitted using the function $R_\text{4g}(T) = a + b_2T^2 + cT^9$, whereas $R_\text{2g}$ and $R_\text{1g}$ data were fitted with $R_\text{ig}(T) = a + b_1T + cT^9$. The best-fit results are presented in Table \ref{Table:rate_recovery}. The observed temperature dependence is consistent with the SLR model, incorporating a T$^{9}$ term associated with Raman processes and linear or quadratic terms representing the combined effect of direct SLR processes. The purely linear dependence observed down to the lowest temperatures for the 1g and 2g recovery rates suggests the absence of any phonon-bottleneck effect, at least along the most efficient relaxation pathways. In contrast, the quadratic temperature dependence exhibited by the 4g recovery rate indicates a strong phonon bottleneck effect in the 4g-3g transition, as further discussed in\,\ref{sec:model1500}. 

At higher temperatures, the Raman term dominates over direct processes. Notably, the Raman coefficient is similar for all recovery rates. In contrast, the linear term b$_1$ in Table\,\ref{Table:rate_recovery} associated with the 1g level population is smaller than that of the 2g level. This difference becomes evident in the absorption spectra in Fig.\,\ref{fig:Figure1}: the two high-energy peaks related to the 1g and 2g levels (see the black arrows) recover at a comparable rate at 2.0\,K but exhibit distinct behaviors at 150\,mK. This suggests that the direct SLR rate differs significantly depending on the level involved.

To gain deeper insight into the complex dynamics of the spin relaxation,  we developed two models to simulate the recovery of the levels population: the first one considers only Raman processes between all possible level pairs, while the second one includes only direct processes. As a result, these models are valid only in the high-temperature regime ($T>$1.7) and low-temperature regime ($T<$1.5), respectively. The models and the data analysis are discussed in Appendices A and B. By fitting the model equations to the appropriate data, we estimated the parameters $c_{ij}$ and $A_{ij}$, which quantify the coupling strength for each Raman and direct process. Our analysis revealed that the coupling strengths $c_{ij}$ associated with Raman processes vary by less than a factor of 3, whereas the $A_{ij}$ parameters of the direct processes span two orders of magnitude depending on the two levels involved. Moreover, the parameter $A_{34}$ and even $A_{12}$ at lower temperature, corresponding to the direct processes between levels with the smallest energy gaps (1-2: 0.528 GHz, 3-4: 0.655 GHz) exhibit a clear reduction with decreasing temperature. Since the coupling strength is expected to be temperature independent, the observed dependence must be attributed to changes in the phonon properties. In particular, this is consistent with the phonon bottleneck phenomenon, where a strong accumulation of phonons at the lowest resonant energy leads to a slowdown in population recovery, resulting in systematically lower extracted values for the $A_{12}$ and $A_{34}$ parameters.

Finally, we observe that the constant term $a$ in Table\,\ref{Table:rate_recovery}, which should be independent of concentration, is larger in the 10\,ppm sample than in the 2\,ppm sample. We attribute this result to an efficient migration of spin polarization throughout the crystal, enabled by FF processes. Since the laser directly polarizes only the Yb ions within the beam path, the significantly faster FF rates in the 10\,ppm sample (see Tab.\,\ref{Table:rate_hole} in the next section) allow the spin perturbation to spread more efficiently toward unperturbed regions of the crystal, and the population within the probed volume to show a faster recovery.

%Now, we should wonder if the recovery at the lowest temperature is dominated by the spontaneous phonon emission or, even in the case of the 2\,ppm sample, is limited by the spin FF effect. 

\begin{table}[ht!]
	\centering
		\caption{Best fit parameters for the temperature dependence of level population recovery time. $a$ is the temperature independent term, b$_1$ and b$_2$ are the coefficients of the direct process, and c is the coefficient of the T$^{9}$ Raman term.}
	\label{Table:rate_recovery}
	\small
		\begin{threeparttable}	
			\begin{tabular}{@{} *1l*3c @{}}   \\
				\toprule
				\textbf{} & a  [10$^{-5}$\,s$^{-1}$] & b$_1$ [10$^{-4}$\,s$^{-1}$K$^{-1}$] \,\,\,& c [10$^{-6}$\,s$^{-1}$K$^{-9}$]   \\ \midrule		
				
				\textbf{1g -  \,\,\,2\,ppm} & 3.5$\pm$0.1 & 2.3$\pm$0.3  & 4.9$\pm$0.5 \\
				\textbf{1g - 10\,ppm} & 19.3$\pm$5.7 & 2.6$\pm$0.2  & 5.1$\pm$0.3 \\
				\textbf{2g -  \,\,\,2\,ppm} & 3.5$\pm$0.2 & 4.3$\pm$0.1  & 4.3$\pm$0.3 \\
    			\textbf{2g - 10\,ppm} & 6.2$\pm$0.7 & 4.4$\pm$0.2  & 5.6$\pm$0.3 \\
       \toprule
				\textbf{} & a [10$^{-5}$\,s$^{-1}$] &  b$_2$ [10$^{-4}$\,s$^{-1}$K$^{-2}$] & c [10$^{-6}$\,s$^{-1}$K$^{-9}$]   \\ \midrule		
                \textbf{4g -  \,\,\,2\,ppm} & 5.6$\pm$1.2 & 2.1$\pm$0.4  & 5.7$\pm$0.4 \\
                \textbf{4g - 10\,ppm} & 10.1$\pm$0.6 & 3.2$\pm$0.1  & 6.1$\pm$0.4 \\
        \bottomrule
			\end{tabular}
		\end{threeparttable}
\end{table}

\subsection{Spectral hole measurements}
\label{sec:SHB}

The spin relaxation was investigated over a wider temperature range using the SHB technique.
Instead of scanning the laser frequency and pumping the entire population of one or more levels, the laser was fixed at a particular frequency. A precise subclass of ions that occupied the same level and shared the same transition frequency within the inhomogeneous absorption linewidth was pumped into an optical state. While these ions occupied the optical level or a hyperfine level of the ground state, different from the initial one after the optical relaxation, a reduced absorption could be observed at the frequency of the initial excitation.

\begin{figure}[h!]
\centering
	\includegraphics[width=0.75\linewidth]{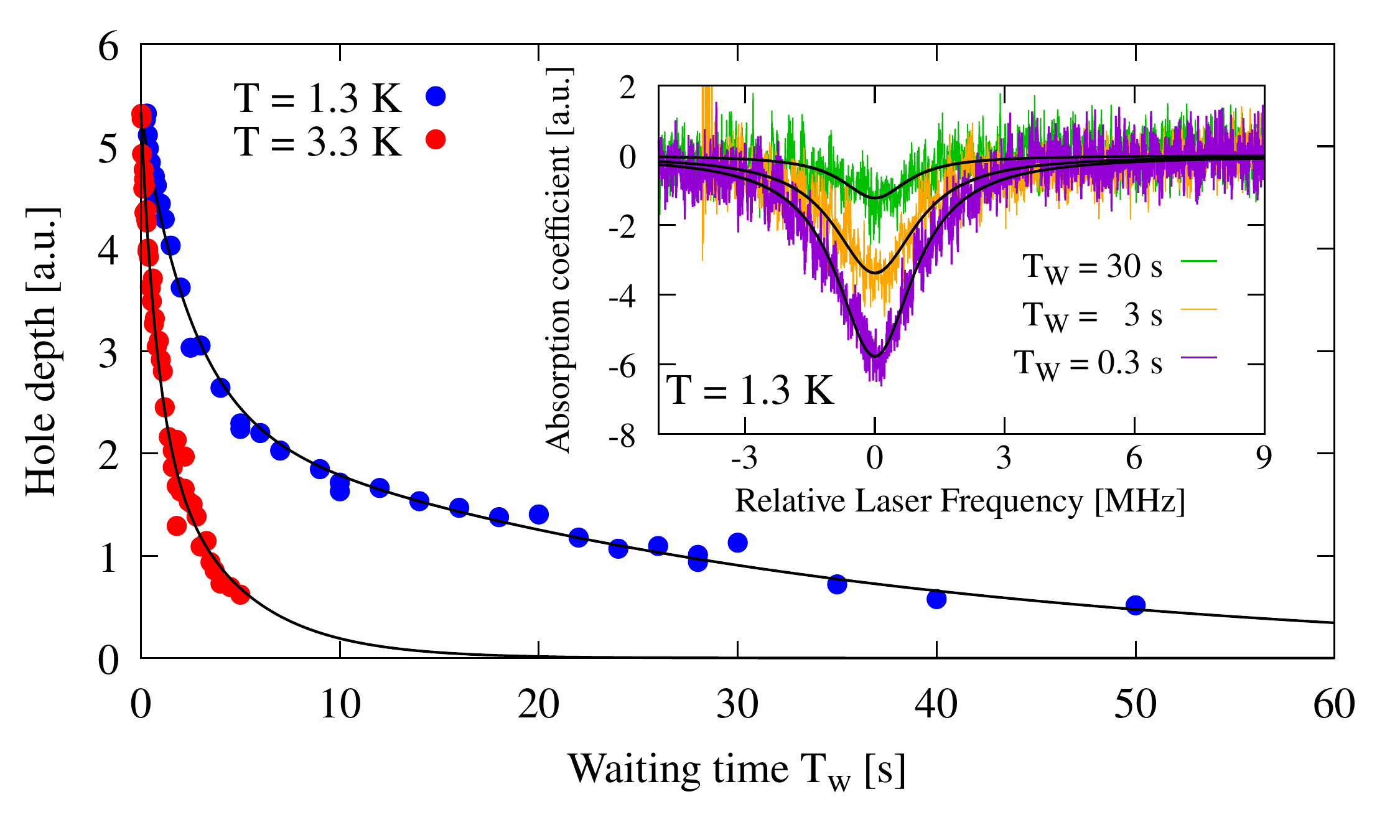}
	\caption{(Inset) Spectral hole profile at different delay times recorded in the absorption spectrum of the 4g-1e transition of the 2\,ppm sample, with Lorentzian fits (black). (Main) Time dependence of the measured hole depth at 1.3\,K and 3.3\,K. The black lines are the curves obtained from the global fit.}
	\label{fig:Figure3}
\end{figure}

In our experiment, we selectively excited a subset of Yb ions at level 4g by tuning the laser to the peak absorption of the 4g-1e transition. After a specific waiting time $T_W$, we scanned the absorption spectrum around the previous excitation frequency and obtained the profile of the burned spectral hole. The burning pulse had a duration of several milliseconds and a power of 10 mW, whereas the probing pulse had a duration of tens of $\mu$s and a power of tens of $\mu$W. This procedure was repeated for different waiting times and representative traces, recorded at a temperature of 1.3\,K for the Yb:YSO 2\,ppm sample are presented in the inset of Fig.\,\ref{fig:Figure3}. Notably, the spectral holes showed no measurable broadening over the entire measurement period.

Since the combination of multiple burning pulses and the FF process could progressively deplete the 4g level population via diffusion enhanced optical pumping (DEOP)\,\cite{Welinski2020}, the entire absorption spectrum was repumped with strong laser pulses before each measurement to reproduce identical initial conditions. The achieved distribution was nearly uniform across the four ground-state levels at each temperature, and we therefore expect to observe the same spin flip-flop dynamics. The main panel of Fig.\,\ref{fig:Figure3} shows with blue circles the spectral hole depth at 1.3\,K as a function of waiting time up to 50\,s in the 2 ppm sample. Two distinct decay components are readily apparent, and qualitatively similar hole decay dynamics were observed between 0.9 K and 2.1 K. Above 2.1\,K, the hole decay accelerates markedly with increasing temperature and the two decay components gradually become more difficult to distinguish. At 3.3\,K (red circles in Fig.\,\ref{fig:Figure3}) they are already comparable. By 4\,K, the decay is effectively a single exponential with a time constant below 1\,s.

The decay of the spectral hole following optical relaxation is governed by two primary mechanisms, which likewise drive the degradation of the AFC. Whereas SLR restores the original population distribution, the FF process redistributes the population deficit of ions from one ground‐state sublevel to all others, thus erasing the hole. Accordingly, the hole decay depends on the total spin relaxation rates $T_\text{ij}=F_\text{ij}+S_\text{ij}$. By neglecting the contribution of the direct process and assuming the symmetric conditions $T_\text{12}=T_\text{34}$, $T_\text{13}=T_\text{24}$,  $T_\text{14}=T_\text{23}$, we simulated the dynamics of the spin relaxation in Sec.\ \ref{sec:model3} and arrived at the following description of the 4g–1e spectral‐hole decay:

\begin{equation}
H(t) = k_1e^{-2t(T_\text{12} + T_\text{13})} + k_2e^{-2t(T_\text{12} + T_\text{14})} + k_3e^{-2t(T_\text{13} + T_\text{14})}.
\label{Eq:hole_eq3}
\end{equation}

The $k_{i}$ coefficients reflect the redistribution of ions that are optically pumped from the 4g level to the 1e level and subsequently relax radiatively into the four hyperfine sublevels of the ground state.

\begin{figure}[h!]
\centering
	\includegraphics[width=0.75\linewidth]{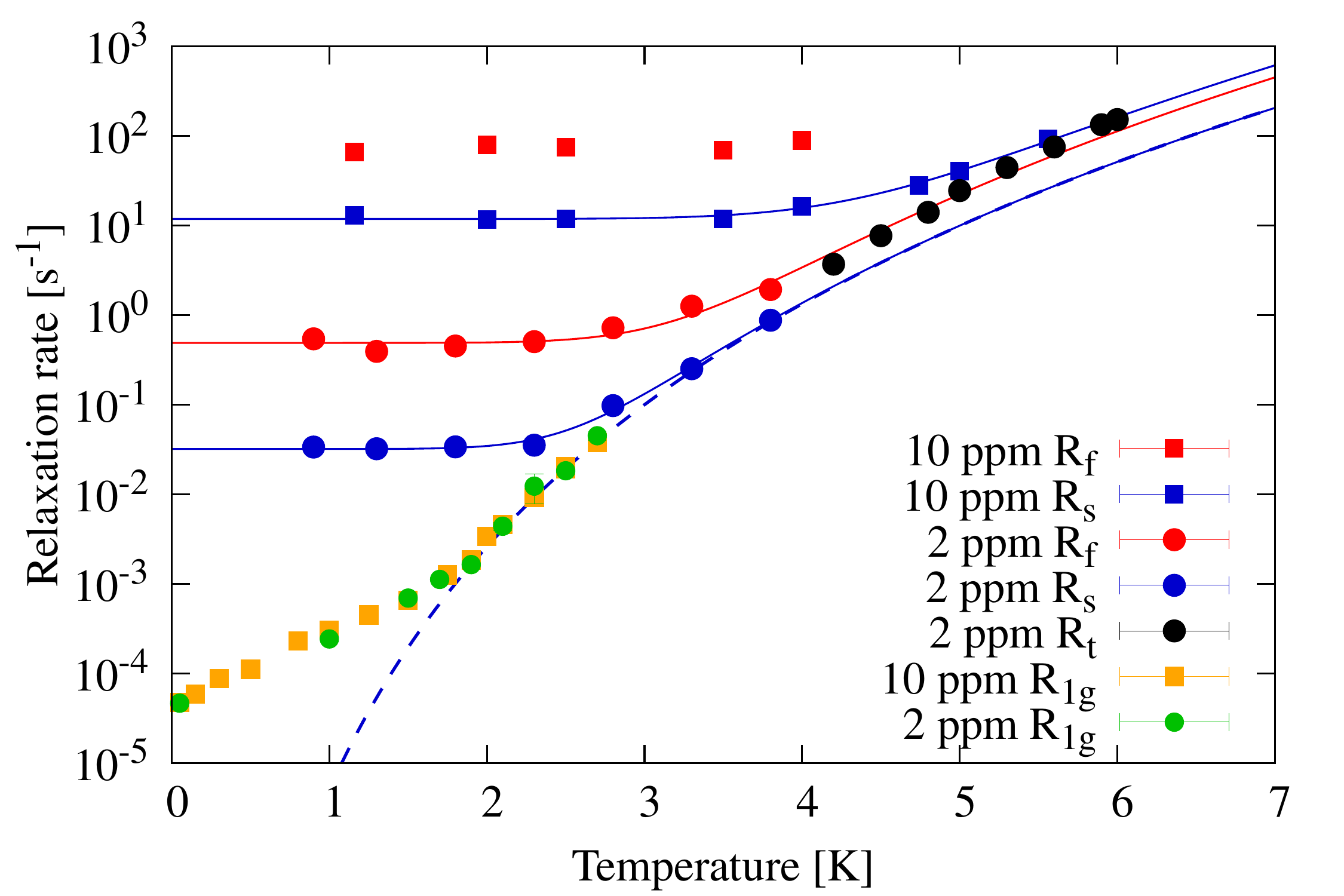}
	\caption{Spectral hole relaxation rates measured by burning the 4g–1e transition in the 10\,ppm (squares) and 2\,ppm (circles) samples. The solid lines represent fits to the data using the function $\bar{a} + \bar{c}T^9$, while the dotted lines indicate the $\bar{c}T^9$ contribution alone. For comparison, population recovery rates R$_{1g}$ associated with the 1g level in both samples are also shown in green and orange.}
	\label{fig:Figure4}
\end{figure}

Below 2.1 K, FF processes dominate the spin relaxation ($T_{ij} \approx F_{ij}$). Using the theoretical hierarchy $F_{12} \gg F_{13} \gg F_{14}$ \cite{Welinski2020}, Eq.,\ref{Eq:hole_eq3} can be approximated as
\begin{equation}
H(t) \approx H(0)\left[(1-k)e^{-tR_f} + ke^{-tR_s}\right],
\label{Eq:hole1}
\end{equation}
where $R_f = 2(T_{12}+T_{13}) \approx 2(T_{12}+T_{14})$ and $R_s = 2(T_{13}+T_{14})$. $H(0)$ denotes the initial hole depth immediately after the burning pulse, and $k$ is a weighting parameter that accounts for the relative amplitudes of the two decay components. 
Over the investigated temperature range, $H(0)$ and $k$ are assumed to remain constant, as identical burning pulses were applied and the excited ions follow the same optical relaxation pathways. Accordingly, all data sets from 0.9\,K to 3.8\,K were fitted simultaneously using a global fit, with $H(0)$ and $k$ treated as global parameters, while $R_f$ and $R_s$ were treated as local parameters specific to each temperature. The global fit yielded a value of $k = 0.45 \pm 0.02$. The obtained local parameters $R_f$ and $R_s$ for the different temperatures are plotted in Fig.\,\ref{fig:Figure4} (red and blue circles, respectively) while the fitted decay curves for representative temperatures of 1.3\,K and 3.3\,K are shown in Fig.\,\ref{fig:Figure3}.

As expected, the hole decay rates remain constant between 0.9\,K and 2.3\,K since they are dominated by the FF processes and the level populations were equalized for all measurements. Above 2.3\,K, Raman processes become significant and the decay rates increase with temperature. The total spin relaxation rate can be thus expressed as $T_{\text{ij}} = F_{\text{ij}} + c_{\text{ij}}T^9$ and the hole decay rates can be approximated as $R_f = 2F_{12} + 2(c_{12}+c_{13})T^9$ and $R_s = 2F_{13} + 2(c_{13}+c_{14})T^9$. Indeed, the experimental data are well described by the simplified expression $R_{\text{s,f}} = \bar{a} + \bar{c}T^9$, as shown in Fig.\,\ref{fig:Figure4}. The fitted values of $\bar{a}$ and $\bar{c}$ for each relaxation component are reported in Table~\ref{Table:rate_hole}. Notably, one of the $\bar{c}$ coefficients is more than twice as large as the other.

Above 4.2\,K, the increasing contribution of Raman processes makes the two hole decay rates progressively closer, such that they can no longer be reliably distinguished by the global fit. Accordingly, we fitted the dataset between 4.5\,K and 6\,K with the function: $H(t) = H(0)\cdot e^{-tR_t}$, with $R_t = 2(T_\text{12}+T_\text{13})\approx 2(T_\text{12}+T_\text{14}) \approx 2(T_\text{12}+T_\text{14})$.
The resulting relaxation rates are plotted with black circles in Fig.\,\ref{fig:Figure4} and also follow a pure $T^9$ dependence, $R_t=\bar{c}T^9$, with $\bar{c}$ matching that of the fast component $R_f$. Above 6 K, the spin‐relaxation rate approaches the optical radiative rate, and our model ceases to be valid.

%\textcolor{blue}{If we consider that one Raman term is larger than the other two and we assume that the derived $1/3T_S$ are more affected by this fast component, then the hole relaxation rates should be closer to $2/3T_S$ rather than $4/3T_S$ and very good compatibility can be achieved.}

\begin{table}[h]
	\centering
		\caption{Best fit parameters for the temperature dependence of hole decay components. $\bar{a}$ represents the temperature independent term due to FF processes and $\bar{c}$ the coefficient of the T$^{9}$ Raman term.}
	\label{Table:rate_hole}
	\small
		\begin{threeparttable}	
			\begin{tabular}{@{} *2l*2c @{}}   \\
				\toprule
				\textbf{sample} & \textbf{} & $\bar{a}$   [s$^{-1}$] &  $\bar{c}$  [10$^{-6}$\,s$^{-1}$K$^{-9}$]   \\ \midrule		
				
				2\,ppm & {$R_f$} & 0.490$\pm$0.044 & 11.1$\pm$2.3  \\
    			2\,ppm & {$R_s$} & 0.032$\pm$0.002 & 5.08$\pm$0.34  \\
                2\,ppm & {$R_T$} & - & 11.8$\pm$0.4  \\
                10\,ppm & {$R_s$} & 11.8$\pm$0.4 & 14.9$\pm$0.8  \\
                 10\,ppm & {$R_f$} & 72.0$\pm$2.8 & - \\
                % 10\,ppm & {$R_\text{ff}$} & 526 & - \\

               % 2\,ppm & {$4/3T_\text{S}$} & - & 24.5$\pm$0.7  \\

        \bottomrule
			\end{tabular}
		\end{threeparttable}
\end{table}

For the 10\,ppm sample, a fast FF component of $\sim$500\,s$^{-1}$ was reported in Ref.\,\cite{Welinski2020, chiossi2024optical}. Extending our analysis to longer delays revealed two additional exponential components, whose rates are included in Fig.\,\ref{fig:Figure4}. Above 4 K, however, the fast component approaches the optical relaxation rate and can no longer be reliably extracted. Even so, the temperature dependence of the slowest component continues to follow the form $\bar{a}+\bar{c}T^9$, with fitted $\bar{c}$ values comparable to those obtained for the 2 ppm sample. Notably, no physically meaningful set of  ($T_\text{12},T_\text{13},T_\text{14}$) values can reproduce all three observed decay rates in the 10\,ppm sample via Eq.\,\eqref{Eq:hole_eq}, suggesting that the true spin‐relaxation dynamics for this sample is more complex than our current model implies (Eqs.\,\ref{Eqs:T}, \ref{sec:model3}).

\section{Discussion and conclusions}

The spectral hole burning and population recovery measurements provide valuable insights into the spin and population relaxation mechanisms of ions at cryogenic and ultra-cryogenic temperatures. While population recovery is governed solely by SLR, the decay of spectral holes encompasses additional contributions from FF processes. Consequently, SLR processes were effectively investigated down to 50\,mK through population recovery measurements, enabling the estimation of the limit of $^{171}$Yb spin lifetime in samples with ultra-low Yb doping. Conversely, the effective spin relaxation rate, predominantly influenced by FF processes at low temperatures, imposes a fundamental limitation on the duty cycle and efficiency of the AFC-based protocol, as determined via the SHB technique. Since the FF processes depends solely on the achieved ion distribution, operating at temperatures below a certain threshold offers no additional advantages. This threshold corresponds to the temperature at which SLR processes become more significant than FF processes. Understanding the rates of FF and SLR processes thus facilitates the selection of optimal ion concentration levels and operating temperatures to achieve a desired AFC degradation time.

The spin flip-flop rates measured in the 2\,ppm Yb:YSO sample are up to three orders of magnitude slower than those observed in the 10\,ppm sample, as detailed in Table \ref{Table:rate_hole}. On the assumption that the spin transition linewidths for the two samples are similar, the observed disparity exceeds the expected scaling based on the concentration ratio, suggesting a more complex dependence. This pronounced concentration dependence may be advantageous for the AFC protocol, despite the underlying mechanisms remaining not fully understood. It implies that significant extensions of spin relaxation times could be achieved with minimal reductions in optical density. Moreover, the FF rates measured under equalized population distributions provide only an upper limit on the AFC relaxation rates. In practical AFC implementations, ions can be concentrated into specific ground-state levels (e.g. 4g and 1g or 4g and 2g), where magnetic coupling is weaker, leading to slower FF rates.

Our findings suggest that the millisecond-scale AFC degradation time observed in 5\,ppm-doped Yb:YSO at 3\,K\,\cite{businger2022non} could be extended to several seconds in a 2\,ppm-doped sample at temperatures below 2.1\,K. If the 3g and 2g levels can be sufficiently depleted, allowing ions in the 4g level to primarily interact with those in the 1g level, degradation times on the order of tens of seconds may be achievable. Furthermore, for concentrations slightly below 2\,ppm, AFC relaxation times exceeding one hour might be possible at temperatures below 1\,K, only limited by direct SLR processes. The phonon-bottleneck effect observed for the 4g–3g transition below 1.5\,K can cause a further reduction of the population relaxation and the AFC degradation rate.

Above 2.1\,K for the 2\,ppm sample and 3.8\,K for the 10\,ppm sample, the spin relaxation rates exhibit a marked increase due to the Raman processes, following a temperature dependence proportional to $T^9$. The extracted Raman coefficients are relatively consistent across the different relaxation pathways. From population recovery measurements, we obtained the following values: $(c_{12}, c_{13}, c_{14}) = (2.47, 1.76, 1.12) \times 10^{-6}$\,s$^{-1}$K$^{-9}$ for the 10\,ppm sample, and $(c_{12}, c_{13}, c_{14}) = (2.92, 1.38, 1.39) \times 10^{-6}$\,s$^{-1}$K$^{-9}$ for the 2\,ppm sample (see Sec.\,\ref{sec:model1700}). These Raman coefficients are in excellent agreement with the results obtained via SHB measurements in the 2\,ppm sample, where $c_{13} + c_{14} = \bar{c}{R_s}/2 = (2.54 \pm 0.17) \times 10^{-6}$\,s$^{-1}$K$^{-9}$ and $c_{12} + c_{14} \approx c_{12} + c_{13} = \bar{c}_{R_f}/2 = (5.5 \pm 1.1) \times 10^{-6}$\,s$^{-1}$K$^{-9}$. The good correspondence is particularly clear in Fig.\,\ref{fig:Figure4}, where the Raman contributions to the population recovery of level 1g and the decay of the spectral hole associated with the slowest transitions overlap significantly.

Above 6\,K, where the spin relaxation rate becomes comparable to the optical radiative rate, the model adopted in this work is no longer valid. In this temperature regime, we had developed a model \cite{chiossi2024optical} that included the dynamics of the optical levels, assuming identical spin relaxation rates between hyperfine levels of the ground and optical levels. The extracted rates still followed a T$^9$ dependence, but the resulting coefficient for each transition, $c = (6.1 \pm 0.2) \times 10^{-6}$\,s$^{-1}$K$^{-9}$, was more than twice the $c_{ij}$ values obtained in this work. This discrepancy likely arises from the larger Raman contributions associated with the optical hyperfine levels, which may lead to an overestimation of the spin relaxation rate when uniform rates are assumed across all levels.

This work underscores that optical probing of population distribution is a powerful tool for investigating spin relaxation processes in rare-earth-doped crystals, especially when multiple ground-state levels are involved. By recording the temporal evolution of the population distribution, either globally across all levels or selectively within specific subclasses of ions, and by developing appropriate models, it is possible to reconstruct the individual relaxation pathways, estimate their associated rates and predict the spin dynamics for arbitrary distributions. This approach proves particularly valuable in systems such as Yb:YSO, where the low symmetry of the crystal field renders first-principles calculations of individual phonon relaxation rates extremely challenging, and the experimental determination of these rates represents the most practical and reliable approach. Estimating these spin flip rates is crucial for quantum technology applications, as they enable inference of spin lifetimes and their influence on the coherent states of surrounding spins.

\section{Acknowledges}
This work was supported by the French National Research Agency through the projects Chorizo (ANR-24-CE47-1190), MIRESPIN (ANR-19-CE47-0011), and the QMemo initiative within the France 2030 program (ANR-22-PETQ-0010). We also acknowledge funding from the European Union’s Horizon 2020 research and innovation programme under Grant Agreement No. 820391 (Square), as well as support from the Swiss National Science Foundation (SNSF, project No. 197168).

\appendix

\section{Estimation of the direct SLR rates from the population recovery}
\label{sec:model1500}
\begin{figure}[h]
	\includegraphics[width=1\linewidth]{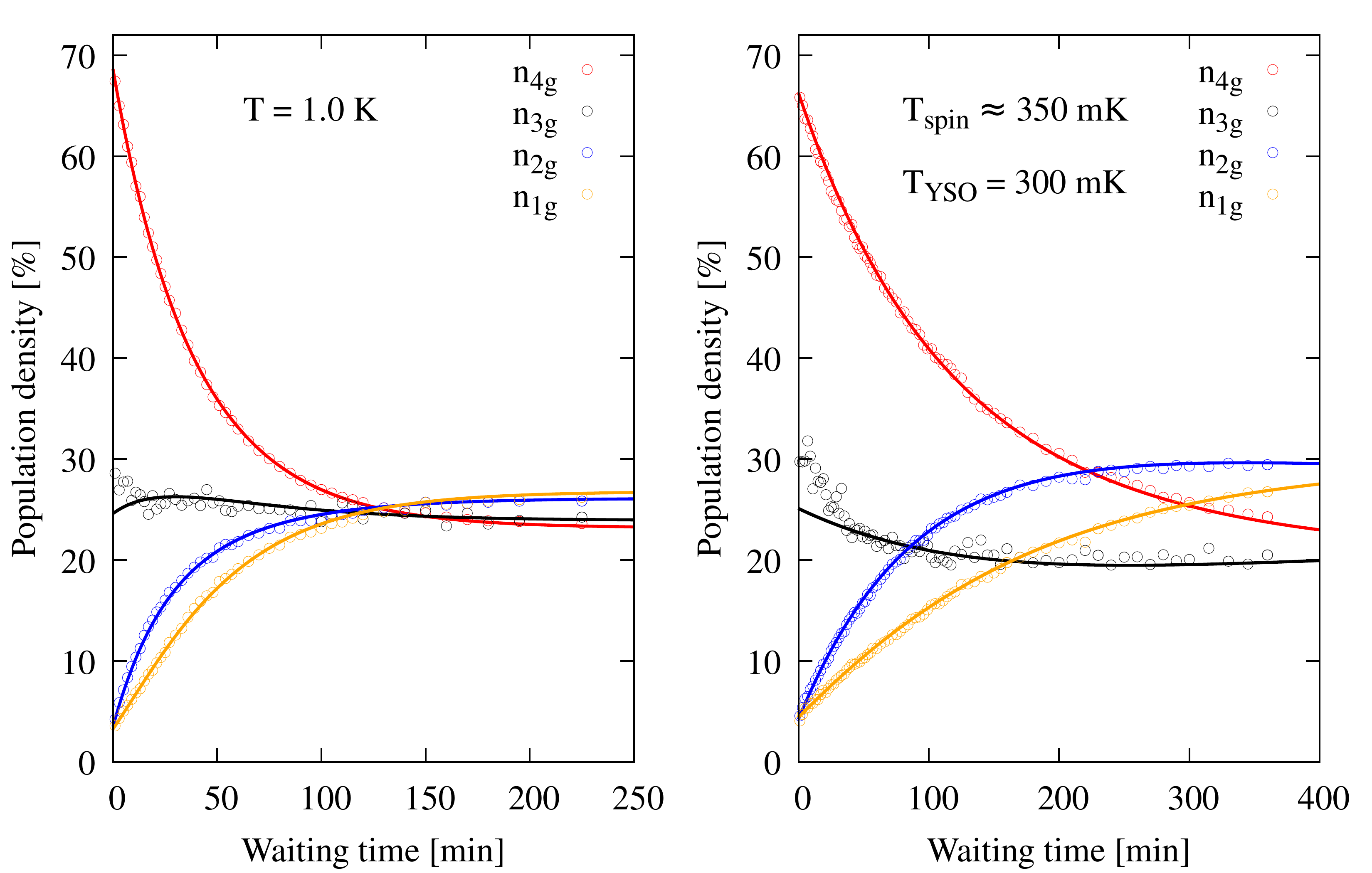}
	\caption{Population densities and relative fits for two crystal temperatures. Even if the data and prediction for the 3g level population density were not used for the fitting procedure, they are shown as well for completeness. It is clear that the 3g population does not vary significantly and is overestimated by the two-Lorentzian fit (Fig.\,\ref{fig:Figure1}c) for the shortest delays.}
	\label{fig:Direct}
\end{figure}

Below 1.5\,K, the Raman contribution becomes negligible compared to the direct process. Spin relaxation for each level is thus governed by direct transitions to and from the other three levels, as described by Eq.\,\ref{Eq:direct_rate}. The spin dynamics are therefore determined by six free parameters, $A_{ij}$, which are too many to allow reliable modeling and accurate parameter estimation. As a first attempt to reduce the number of free parameters, we assumed the following symmetry conditions: $A_{12} = A_{34}$, $A_{13} = A_{24}$, and $A_{14} = A_{23}$. However, using only three free parameters, we were not able to satisfactorily reproduce the population recovery dynamics. A good agreement could only be achieved by relaxing the condition $A_{12} = A_{34}$ and modeling the population evolution using the following system of equations:

\vspace{0.5cm}
\noindent
\resizebox{1\textwidth}{!}{$
%    \left \{
\begin{array}{rcl}
&\dfrac{dn_\text{1g}}{dt} = -\dfrac{(\hbar \omega_\text{12})^3A_\text{12}(n_\text{1g}-n_\text{2g}k_\text{12})}{k_\text{12}-1}-\dfrac{(\hbar \omega_\text{13})^3A_\text{24}(n_\text{1g}-n_\text{3g}k_\text{13})}{k_\text{13}-1}-\dfrac{(\hbar \omega_\text{14})^3A_\text{14}(n_\text{1g}-n_\text{4g}k_\text{14})}{k_\text{14}-1} \\  \\
&\dfrac{dn_\text{2g}}{dt} = -\dfrac{(\hbar \omega_\text{12})^3A_\text{12}(n_\text{2g}k_\text{12}-n_\text{1g})}{k_\text{12}-1}-\dfrac{(\hbar \omega_\text{24})^3A_\text{24}(n_\text{2g}-n_\text{4g}k_\text{24})}{k_\text{24}-1}-\dfrac{(\hbar \omega_\text{23})^3A_\text{14}(n_\text{2g}-n_\text{3g}k_\text{23})}{k_\text{23}-1} \\ \\
&\dfrac{dn_\text{3g}}{dt} = -\dfrac{(\hbar \omega_\text{34})^3A_\text{34}(n_\text{3g}-n_\text{4g}k_\text{34})}{k_\text{34}-1}-\dfrac{(\hbar \omega_\text{13})^3A_\text{24}(n_\text{3g}k_\text{13}-n_\text{1g})}{k_\text{13}-1}-\dfrac{(\hbar \omega_\text{23})^3A_\text{14}(n_\text{3g}k_\text{23}-n_\text{2g})}{k_\text{23}-1} \\ \\
&\dfrac{dn_\text{4g}}{dt} = -\dfrac{(\hbar \omega_\text{34})^3A_\text{34}(n_\text{4g}k_\text{34}-n_\text{3g})}{k_\text{34}-1}-\dfrac{(\hbar \omega_\text{24})^3A_\text{24}(n_\text{4g}k_\text{24}-n_\text{2g})}{k_\text{24}-1}-\dfrac{(\hbar \omega_\text{14})^3A_\text{14}(n_\text{4g}k_\text{14}-n_\text{1g})}{k_\text{14}-1} \\ \\
\end{array}
%\right.
$
}
with $k_\text{ij} = \exp(\hbar \omega_\text{ij}/k_bT)$. $A_\text{34}$, $A_\text{12}$, $A_\text{24}$ and $A_\text{14}$ are the only free parameters. As boundary conditions, we set each initial population to the one measured at the shortest delay. Unfortunately, the analytical solution is too long to be shown here, but we can mention that the solution for each population at a certain temperature and $A_\text{ij}$ values is of the following form:
\begin{equation}
    n_\text{ig}(t) = n_\text{ig}^{eq} + a_\text{1i}e^{-tR_1} + a_\text{2i}e^{-tR_2} + a_\text{3i}e^{-tR_3}.
\end{equation}
In other words, the population density thermalization is characterized by three exponential decay, whose rate factors are the same for each level. %The fact that the decay could be approximated by a single exponential function indicates that $R_1$, $R_2$, and $R_3$ values for a certain temperature are not too different. 
The $A_\text{ij}$ parameters have been determined by minimizing the function:
\begin{equation}
    \chi^2(A_\text{12},A_\text{13},A_\text{14},A_\text{34})  = \sum_i \left({n_\text{1g}^\text{th}-n_\text{1g}^\text{exp}} \right)^2 + \sum_i \left({n_\text{2g}^\text{th}-n_\text{2g}^\text{exp}} \right)^2 + \sum_i \left({n_\text{4g}^\text{th}-n_\text{4g}^\text{exp}}\right)^2
\end{equation}
in which the index $i$ runs over the experimental data taken at different delay times. Note that the $n_\text{ig}^{eq}$ values are solely determined by the fixed temperature as the population at equilibrium must follow the Boltzmann distribution in the employed model. For the dataset collected below 500\,mK, the crystal temperature no longer matches the spin temperature and cannot be used for our model. Therefore, we used an effective spin temperature determined by the population distribution measured at long delay times. The fitted curves approximate well the population decay for all datasets between 150\,mK (spin temperature = 240\,mK) and 1.75\,K, as shown for two examples in Fig.\,\ref{fig:Direct}. The failure to accurately fit the dataset at the lowest temperature (50\,mK) might be caused by more transitions becoming phonon bottlenecked, invalidating the $A_\text{13}=A_\text{24}$ and $A_\text{14}=A_\text{23}$ conditions. Furthermore,  the effect of the radiation emitted by the 4\,K thermal shield and the effect of spin distribution migration over the crystal have become increasingly relevant for lower dilution base temperatures and could also play an important role.

\begin{figure}[h]
\centering
	\includegraphics[width=0.7\linewidth]{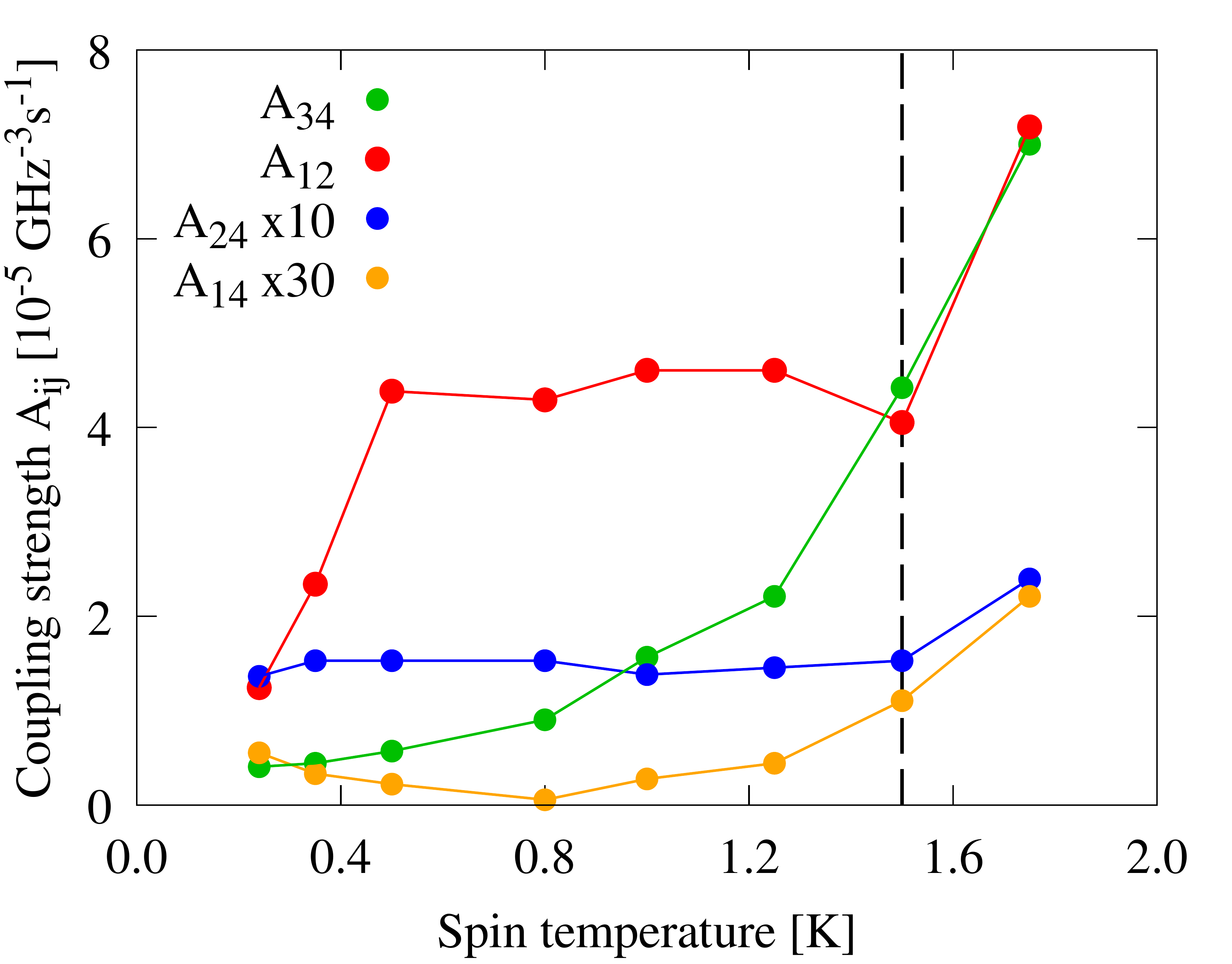}
	\caption{Best-fit coupling strength $A_\text{ij}$ parameters for different effective spin temperatures. The relative errors are below 10\%. The colored lines are drawn to guide eyes. }
	\label{fig:Aij}
\end{figure}

The best-fit $A_{ij}$ parameters extracted from the population recovery data at different temperatures are shown in Fig.\,\ref{fig:Aij}. The results reveal the following hierarchy: $A_{12} \geq A_{34} \gg A_{13} = A_{24} \gg A_{14} = A_{23}$, closely resembling the relation calculated for spin flip-flop processes. To gain further insight into the effective coupling strengths between levels, we computed the direct downward relaxation rates, defined as $S_{ij} = A_{ji}(\hbar \omega_{ji})^3/[\exp(\hbar \omega_{ji}/k_\mathrm{B}T) - 1]$, for three representative temperatures. The resulting values are reported in Table\,\ref{tab:rate_direct}. Among them, $S_{14}$ and $S_{23}$ are significantly smaller than the other relaxation rates. The remaining transitions exhibit similar effective rates, as a weaker coupling strength typically compensates a higher transition frequency. %For instance, $\omega_\text{12}$ and $\omega_\text{34}$ are at least three times smaller than $\omega_\text{13}$ and $\omega_\text{24}$. 

\begin{table}[ht!]
	\centering
		\caption{Spontaneous phonon emission rates $S_\text{ij} = (\hbar\omega_\text{ij})^3A_\text{ij}/(\exp(\hbar \omega_\text{ij}/k_bT)-1)$ in 10$^{-5}$\,s$^{-1}$. The relative transition energy is reported in the last row and the values are in GHz.}
	\label{tab:rate_direct}
	\small
		\begin{threeparttable}	
			\begin{tabular}{@{} *1c*6c @{}}   \\
				\toprule
				\textbf{T} & \textbf{$S_\text{12}$} & \textbf{$S_\text{13}$} & \textbf{$S_\text{14}$}  & \textbf{$S_\text{23}$} & \textbf{$S_\text{24}$} & \textbf{$S_\text{34}$} \\
    \midrule
                1.5 K & 38.2 & 25.8 & 10.0  & 3.79 & 28.6 & 58.7 \\
				1.0 K & 26.4 & 15.3 & 1.63  & 0.62 & 16.9 & 13.8 \\
                0.5 K & 12.2 & 7.97 & 0.60 & 0.24 & 8.79 & 2.47 \\	
                \midrule
                 $\mathbf{\hbar \omega_\text{ij}}$ & 0.53 & 2.37 & 3.03 & 1.84 & 2.50 & 0.66 \\
        \bottomrule
			\end{tabular}
		\end{threeparttable}
\end{table}

%, with values varying between $1.8\,10^{-8}$ and $1.8\,10^{-7}$ in the (0.2-1.4)\,K temperature range. 

%Finding the coupling parameters $A_\text{ij}$ identical for all temperatures investigated would support the validity of the proposed model, even though particular deviations 

%if the phonon bottleneck effect becomes relevant for a certain transition for decreasing temperatures, the relaxation rate will be smaller than expected and the fit procedure will thus give a smaller $A_\text{ij}$ parameter for that transition. On the other hand, for temperatures above 1.5\,K, all spin relaxation rates are significantly enhanced by the Raman processes
%and the fit is expected to give larger coupling strength to take into account the additional mechanism.

The coupling strength factors $A_{ij}$ are expected to be temperature-independent, and indeed both $A_{12}$ and $A_{24} = A_{13}$ remain constant within the (0.5–1.5)\,K temperature range. The increase of all $A_{ij}$ parameters above 1.5\,K can be attributed to the growing influence of Raman processes on spin relaxation, while the applied model accounts only for direct processes. A decreasing trend in $A_{34}$ and also in $A_{12}$ below 500\,mK is observed. However, this behavior does not necessarily invalidate our results. At very low temperatures, the phonon bottleneck effect can become increasingly significant, suppressing the relaxation rate more than predicted by the Debye model. In such cases, the fitting procedure can still recover the correct spin relaxation rates by compensating through decreasing the corresponding $A_{ij}$ factor. This effect is particularly evident for the 3g–4g and 1g–2g transitions, which involve the lowest energy splittings and, consequently, the fewest resonant phonon modes in the crystal. Interestingly, the bottleneck for the 1g–2g transition appears only below 500\,mK, despite its energy being lower than that of the 3g–4g transition. This can be understood by considering the initial population distribution: the optical pumping significantly populates the 4g level while depleting the 1g and 2g levels. During spin thermalization, fewer phonons matching the 1g–2g transition energy are generated compared to the 3g–4g case, reducing the impact of the phonon bottleneck effect. It is interesting to note that in Fig.\,\ref{fig:Aij} starting from 1.5\,K, even $A_\text{12}=A_\text{34}$ holds, suggesting that the discrepancy at lower temperature is indeed originating from the phonon bottleneck effect.

No bottleneck effect is expected for the 1g–4g and 2g–3g transitions down to 150\,mK due to the large energy separation involved. The irregular behavior of $A_{14}$ is likely due to higher uncertainties in their determination. This parameter, is the smallest among all $A_{ij}$, and indeed Raman contributions become non-negligible already at 1.5\,K.

\section{Estimation of the Raman SLR rates from the population recovery}
\label{sec:model1700}

For temperatures higher than 1.7\,K, the population recovery rates $R_\text{ig}$ are dominated by the Raman processes, featuring the specific $T^9$ temperature dependence. Contrary to the direct process, the Raman process rate is expected to be independent of the hyperfine transition energy. In addition, at these temperatures, the populations of the hyperfine levels at thermal equilibrium differ by less than 8\% and we can thus reasonably assume that the upward spin-flip rate equals the downward spin-flip rate: $S_\text{23}=S_\text{14}$ and $S_\text{13}=S_\text{24}$.  We add the condition $S_\text{12}=S_\text{34}$ for the high-temperature model. The system of differential equations approximating the population-level dynamics finally reads:

\begin{equation}
    \left \{
\begin{array}{rcl}
&\dfrac{dn_\text{1g}}{dt} = -S_\text{12}(n_\text{1g}-n_\text{2g})-S_\text{13}(n_\text{1g}-n_\text{3g})-S_\text{14}(n_\text{1g}-n_\text{4g}) \\  \\
&\dfrac{dn_\text{2g}}{dt} = -S_\text{12}(n_\text{2g}-n_\text{1g})-S_\text{13}(n_\text{2g}-n_\text{4g})-S_\text{14}(n_\text{2g}-n_\text{3g}) \\ \\
&\dfrac{dn_\text{3g}}{dt} = -S_\text{12}(n_\text{3g}-n_\text{4g})-S_\text{13}(n_\text{3g}-n_\text{1g})-S_\text{14}(n_\text{3g}-n_\text{2g}) \\ \\
&\dfrac{dn_\text{4g}}{dt} = -S_\text{12}(n_\text{4g}-n_\text{3g})-S_\text{13}(n_\text{4g}-n_\text{2g})-S_\text{14}(n_\text{4g}-n_\text{1g}) \\ \\
\end{array}
\right.
\label{Eqs:system_1700}
\end{equation}

The solution of this system of ordinary differential equations is:
\begin{equation}
    \left \{
\begin{array}{rcl}
&n_\text{1g} = 0.25 + N_\text{14}^{23}e^{-2t(S_\text{12} + S_\text{13})} - N_\text{24}^{13}e^{-2t(S_\text{12} + S_\text{14})} - N_\text{34}^{12}e^{-2t(S_\text{13} + S_\text{14})} \\ \\
&n_\text{2g} =  0.25 - N_\text{14}^{23}e^{-2t(S_\text{12} + S_\text{13})} + N_\text{24}^{13}e^{-2t(S_\text{12} + S_\text{14})} - N_\text{34}^{12}e^{-2t(S_\text{13} + S_\text{14})} \\ \\
&n_\text{3g} =  0.25 - N_\text{14}^{23}e^{-2t(S_\text{12} + S_\text{13})} - N_\text{24}^{13}e^{-2t(S_\text{12} + S_\text{14})} - N_\text{34}^{12}e^{-2t(S_\text{13} + S_\text{14})}  \\ \\
&n_\text{4g} = 0.25 + N_\text{14}^{23}e^{-2t(S_\text{12} + S_\text{13})} + N_\text{24}^{13}e^{-2t(S_\text{12} + S_\text{14})} + N_\text{34}^{12}e^{-2t(S_\text{13} + S_\text{14})} \\ 
\end{array}
\right.
\label{Eqs:sol_recovery}
\end{equation}
where $N_\text{n4}^{ij} = (N_\text{4g}+N_\text{ng}-N_\text{ig}-N_\text{jg})/4$ and $N_\text{1g}+N_\text{2g}+N_\text{3g}+N_\text{4g}=1$. We notice that the solution of each level population is again a sum of four terms. A constant representing the population at equilibrium and three terms composed of a factor $N_\text{nm}^{ij}$, which is an algebraic sum of the initial level populations, multiplied by a time-dependent exponential factor containing the sum of two $S_\text{1j}$ SLR rates. %It is worth mentioning that the initialization laser pulses drove more than 65\% of the Yb ions into the 4g level (see Fig.a), leaving only $\sim$5\% in the level 1g and 2g. All $N_\text{n4}^{ij}$ are therefore positive and constrained between $N_\text{24}^{13}\geq0.1$ and $N_\text{34}^{12}\leq0.2$. Therefore, the recovery of the 4g population should be characterized by three exponential decays with similar positive amplitude. However, only one exponential decay is sufficient to approximate the recovery data for each temperature well, thereby indicating that the three exponential terms must be very similar, namely $S_\text{12}\approx S_\text{13}\approx S_\text{14}\approx R_\text{4g}/4$. From the latter and the fact $R_\text{4g}$ scale with $T^9$, we can deduce that each $S_\text{ij}$ is predominantly represented by the Raman term $c_\text{ij}T^9$ for $T>2\,$K and the Raman coefficient $c_\text{ij}$ are not too different from one level pair to another. 

\begin{figure}[h!]
	\includegraphics[width=1\linewidth]{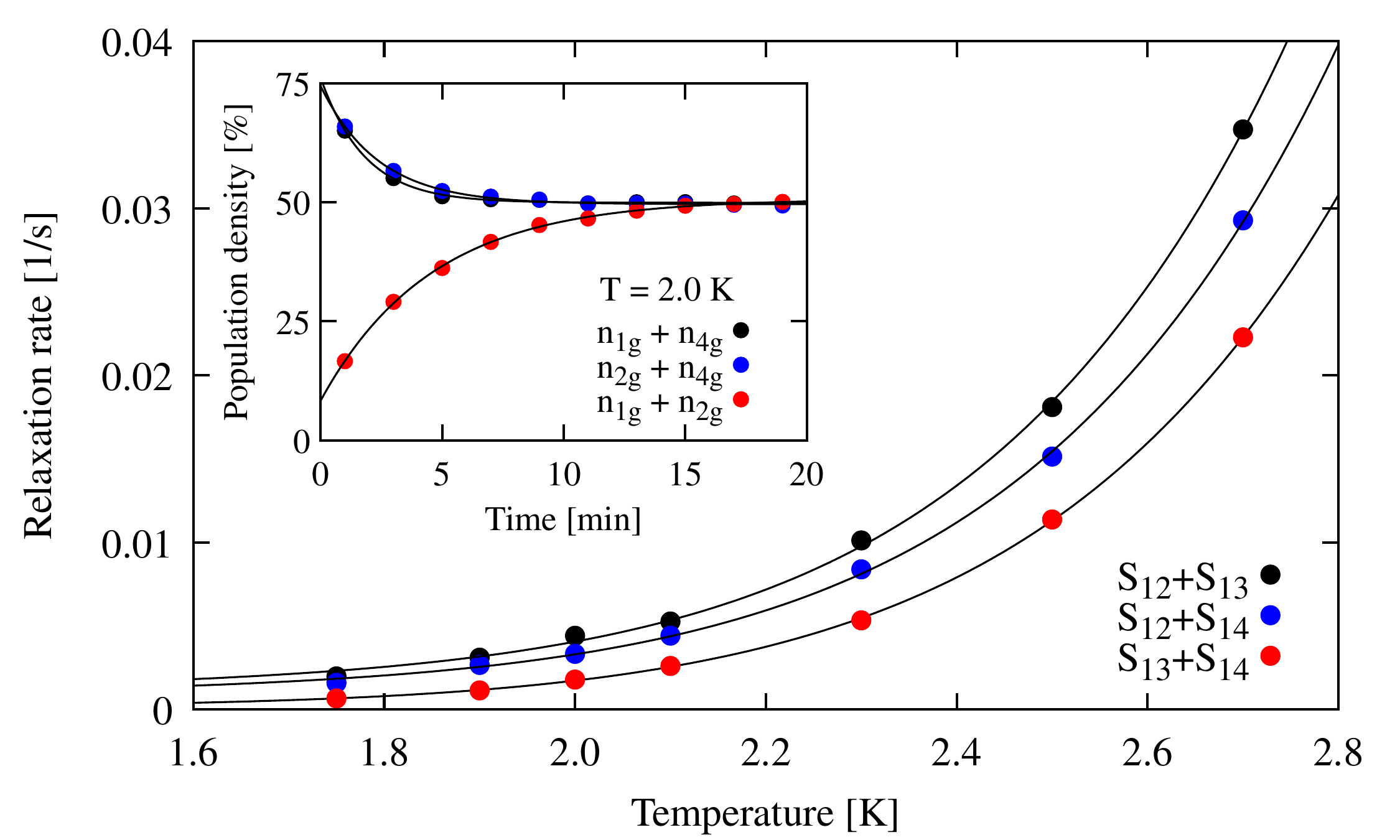}
	\caption{(Inset) Evolution of the sum of two population-level densities. The black lines are single-exponential fit. Temperature dependence of the three two-level population relaxation rates. The black lines are the best-fit results using the function $bT+cT^9$. }
	\label{fig:refined}
\end{figure}

An estimation of these sums can be extrapolated in the following way. Let us consider the sum of the following population densities:

\begin{equation}
    \left \{
\begin{array}{rcl}
&n_\text{1g}+n_\text{2g} = 0.5  - 2N_\text{34}^{12}e^{-2t(S_\text{13} + S_\text{14})} \\ \\
&n_\text{2g}+n_\text{4g}=  0.5 + 2N_\text{24}^{13}e^{-2t(S_\text{12} + S_\text{14})}  \\ \\
&n_\text{1g}+n_\text{4g} = 0.5 + 2N_\text{14}^{23}e^{-2t(S_\text{12} + S_\text{13})}  \\ 
\end{array}
\right.
\label{Eqs:rido}
\end{equation}
where the right-hand side terms have been derived from Eqs.\,\ref{Eqs:sol_recovery}. Each equation in Eqs.\,\ref{Eqs:rido} thus depends on only one combination of $S_\text{1j}+S_\text{1k}$, which can be estimated through an exponential fit for any temperature. An example is shown in the inset of Fig.\,\ref{fig:refined}.

The obtained $S_\text{1j}+S_\text{1k}$ values are fitted with the function $bT+cT^9$ to include both the contribution of the direct and Raman processes for both 1$\rightarrow$j and 1$\rightarrow$k transitions (Fig.\,\ref{fig:refined}). These $c$ values represent the sum of the Raman coefficients $c_\text{1j}+c_\text{1k}$ of the two transitions and having estimated the sums for all three $j,k$ combinations, we can calculate all three $c_\text{1j}$ terms. We obtain $(c_\text{12},c_\text{13},c_\text{14})=(2.47,1.76,1.12)\cdot10^{-6}\,$Hz/K$^9$ for the 10\,ppm sample, with relative errors below 1\%, and $(c_\text{12},c_\text{13},c_\text{14})=(2.92,1.38,1.39)\cdot10^{-6}\,$Hz/K$^9$ for the 2\,ppm sample, with relative errors around 10\,\%.

It is worth noticing that the obtained $c_\text{ij}$ parameters are quite close and so are the exponential factors that appear in the population equations in Eqs\,\ref{Eqs:sol_recovery}. It is therefore no wonder that the individual population decay could be well-fitted with a single exponential function.

\section{Model for the spectral holes}
\label{sec:model3}
In describing the evolution of the spectral hole depth, we focus on the fraction $n_{4g}'$ of Yb ions in the 4g level whose 4g–1e transition frequency falls within the linewidth of the spectral hole. Additionally, we consider ions initially in the 1g, 2g, and 3g levels that, once transferred to the 4g level, also possess transition frequencies falling within the same spectral range. It is important to note that the holes created, with linewidths on the order of 1\,MHz in the 2\,ppm doped crystal, do not exhibit noticeable broadening during the measurement time. This indicates that spectral diffusion is much smaller than the intrinsic spectral hole width. As a result, the subset of Yb ions contributing to absorption at the spectral hole frequencies when occupying the 4g level is well-defined, namely it remains unchanged throughout the entire measurement period. We can therefore restrict our model for simulating the hole decay dynamics to the population distribution of these specific ions, denoted as $n_{1g}'$, $n_{2g}'$, $n_{3g}'$, and $n_{4g}'$.

The rate $T_{ij}$ of the process transferring an ion from the $n_{ig}'$ level to the $n_{jg}'$ level depends on both interactions with the crystal lattice and interactions with the entire Yb ion population.  Under our experimental conditions, this rate per ion remains constant over time, is independent of the evolution of the $n_{ig}'$ distribution, and is equal to the per-ion rate $T_{ji}$ of the inverse process. Indeed, the SLR contribution depends solely on temperature, while the FF process is not restricted to the subdistribution $n_{i}'$, but involves the entire Yb ion population. The latter is equalized at the beginning of each measurement through the initialization pulses and can be considered constant during the spectral hole evolution, given that the subpopulation $n_{i}'$ represents only a very small fraction ($<$1\%) of the total ion population.

The rate equation for the levels subclass population density ($n'_\text{1g},\,n'_\text{2g},\,n'_\text{3g},\,n'_\text{4g}$) can be thus expressed in a similar form of Eqs.\,\ref{Eqs:system_1700}:

\begin{equation}
    \left \{
\begin{array}{rcl}
&\dfrac{dn'_\text{1g}}{dt} = -T_\text{12}(n'_\text{1g}-n'_\text{2g})-T_\text{13}(n'_\text{1g}-n'_\text{3g})-T_\text{14}(n'_\text{1g}-n'_\text{4g}) \\  \\
&\dfrac{dn'_\text{2g}}{dt} = -T_\text{12}(n'_\text{2g}-n'_\text{1g})-T_\text{13}(n'_\text{2g}-n'_\text{4g})-T_\text{14}(n'_\text{2g}-n'_\text{3g}) \\ \\
&\dfrac{dn'_\text{3g}}{dt} = -T_\text{12}(n'_\text{3g}-n'_\text{4g})-T_\text{13}(n'_\text{3g}-n'_\text{2g})-T_\text{14}(n'_\text{3g}-n'_\text{2g}) \\ \\
&\dfrac{dn'_\text{4g}}{dt} = -T_\text{12}(n'_\text{4g}-n'_\text{3g})-T_\text{13}(n'_\text{4g}-n'_\text{2g})-T_\text{14}(n'_\text{4g}-n'_\text{1g}) \\ \\
\end{array}
\right.
\label{Eqs:T}
\end{equation}
The system of equations has the same form as Eqs.\,\ref{Eqs:system_1700}, so that the solution for $n_{i,4g}$ is given by: 
 \begin{equation}
n_\text{4g}(t) = 0.25 - k_1e^{-2t(T_\text{12} + T_\text{13})} - k_2 e^{-2t(T_\text{12} + T_\text{14})} - k_3 e^{-2t(T_\text{13} + T_\text{14})}.
\end{equation}
The parameters $k_i$ depend on the fraction of ions $N'$ pumped from the 4g level to the 1e level, and on the subsequent relaxation of these ions among the four ground-state hyperfine levels. Consequently, the evolution of the hole depth (or area) $H(t)$, which is proportional to $0.25-n_\text{4g}(t)$, is thus given by:
\begin{equation}
H(t) =  k_1e^{-2t(T_\text{12} + T_\text{13})} + k_2 e^{-2t(T_\text{12} + T_\text{14})} + k_3 e^{-2t(T_\text{13} + T_\text{14})}.
\label{Eq:hole_eq}
\end{equation}

%$The initial conditions of the $n_{ig}'$ distribution depend on the fraction of ions $N'$ pumped from the 4g level to the 1e level, and on how these ions subsequently relax among the four ground-state hyperfine levels. %As a first approximation, experimentally validated in Ref.\,\cite{chiossi2024optical}, we can use the branching ratios ($(\beta_\text{1e,1g},\,\beta_\text{1e,2g},\,\beta_\text{1e,3g},\,\beta_\text{1e,4g})=(0.15,\,0.06,\,0.07,\,0.72)$ determined in Ref.\,\cite{Welinski2020} for the optical relaxation along the D1 axis. Under these assumptions, the initial population distribution density immediately after optical relaxation can be expressed as:
%\begin{equation}
%(0.25+N\beta_\text{1e,1g},\,0.25+N\beta_\text{1e,2g},\,0.25+N\beta_\text{1e,3g},\,0.25-N(1-\beta_\text{1e,4g}))
%\end{equation}
%with 0.25 the subclass population density assumed for each hyperfine level at equilibrium. The solution for the subclass population density in 4g level becomes:

\printbibliography

\end{document}